\documentclass[compsoc, conference, a4paper, 10pt, times]{IEEEtran}

\usepackage{cite}
\usepackage{amsmath,amssymb,amsfonts}
\usepackage{algorithmic}
\usepackage{graphicx}
\usepackage{textcomp}
\usepackage{xcolor}

\usepackage{url}
\usepackage{xspace}
\usepackage[binary-units=true]{siunitx}
\DeclareSIUnit \voltampere { VA } %
\DeclareSIUnit \var { var } %

\usepackage{flushend} %

\usepackage{footmisc}
\makeatletter
\newcommand\thefontsize{The current font size is: \f@size pt}
\makeatother

\usepackage{multirow}
\usepackage{paralist}
\usepackage[shortlabels]{enumitem}
\usepackage[normalem]{ulem} %
\usepackage{acronym}
\usepackage{tabularx}
\usepackage{booktabs}
\renewcommand{\arraystretch}{0.9}

\usepackage{dblfloatfix}

\acrodef{ac}[AC]{alternating current}
\acrodef{arp}[ARP]{Address Resolution Protocol}
\acrodef{des}[DES]{discrete-event simulation}
\acrodef{dos}[DoS]{Denial-of-Service}
\acrodef{fms}[FMS]{Facility Monitoring System}
\acrodef{hil}[HIL]{hardware-in-the-loop}
\acrodef{ict}[ICT]{Information and Communications Technology}
\acrodef{ids}[IDS]{Intrusion Detection System}
\acrodef{ied}[IED]{Intelligent Electronic Device}
\acrodef{ioa}[IOA]{information object address}
\acrodef{mact}[MACT]{maximum admissible conductor temperature}
\acrodef{mtu}[MTU]{Master Terminal Unit}
\acrodef{mitm}[MitM]{machine-in-the-middle}
\acrodef{pv}[PV]{photovoltaic}
\acrodef{rtu}[RTU]{Remote Terminal Unit}
\acrodef{se}[SE]{state estimation}
\acrodef{vcc}[VCC]{Virtual Control Center}

\colorlet{RED}{red}

\newcommand{\NAME}{\textsc{Wattson}\xspace}

\usepackage{wasysym}
\usepackage[notextcomp]{stix}
\usepackage{pifont}

\usepackage{tikz}
\usetikzlibrary{calc}
\newcommand{\sw}{0.8em}
\newcommand{\squareF}[1]{%
\tikz{%
\fill[black] rectangle(\sw,#1*\sw);
\draw[line width=0.3pt] rectangle(\sw,\sw);%
}}%

\newcommand{\cmark}{\ding{51}}

\newcommand{\rateYes}{\cmark}
\newcommand{\rateNo}{}
\newcommand{\rate}[1]{{%
\ifnum#1=0 \pie{100}%
	\else\ifnum#1=1 \squareF{0}%
	\else\ifnum#1=2 \squareF{0}%
	\else\ifnum#1=3 \squareF{0.33}%
	\else\ifnum#1=4 \squareF{0.66}%
	\else\ifnum#1=5 \squareF{1}%
	\fi\fi\fi\fi\fi\fi%
}}

\newcommand{\fns}{\makebox(0,0)[lb]{$^\text{\small*}$}}
\newcommand{\rateU}[1]{{%
\ifthenelse{\equal{#1}{\string ?}}
  {?}
  {\phantom{\fns}\rate{#1}\fns}%
}}

\begin{document}

\title{Comprehensively Analyzing the Impact of Cyberattacks on Power Grids}

\author{\IEEEauthorblockN{Lennart Bader}
\IEEEauthorblockA{%
\textit{Fraunhofer FKIE \&}\\
\textit{RWTH Aachen University}\\
lennart.bader@fkie.fraunhofer.de}
\\
\IEEEauthorblockN{Ömer Sen}
\IEEEauthorblockA{\textit{Fraunhofer FIT \&}\\
\textit{RWTH Aachen University}\\
o.sen@iaew.rwth-aachen.de}
\\
\IEEEauthorblockN{Julian Filter}
\IEEEauthorblockA{\textit{RWTH Aachen University}\\
hallo@julianfilter.de}
\and
\IEEEauthorblockN{Martin Serror}
\IEEEauthorblockA{%
\textit{Fraunhofer FKIE}\\
martin.serror@fkie.fraunhofer.de}
\\
\\
\IEEEauthorblockN{Dennis van der Velde}
\IEEEauthorblockA{\textit{Fraunhofer FIT}\\
dennis.van.der.velde@fit.fraunhofer.de}
\\
\\
\IEEEauthorblockN{Elmar Padilla}
\IEEEauthorblockA{%
\textit{Fraunhofer FKIE}\\
elmar.padilla@fkie.fraunhofer.de}
\and
\IEEEauthorblockN{Olav Lamberts}
\IEEEauthorblockA{%
\textit{Fraunhofer FKIE \&}\\
\textit{RWTH Aachen University}\\
olav.lamberts@rwth-aachen.de}
\\
\IEEEauthorblockN{Immanuel Hacker}
\IEEEauthorblockA{\textit{Fraunhofer FIT \&}\\
\textit{RWTH Aachen University}\\
i.hacker@iaew.rwth-aachen.de}
\\
\IEEEauthorblockN{Martin Henze}
\IEEEauthorblockA{%
\textit{RWTH Aachen University}\\
\textit{\& Fraunhofer FKIE}\\
henze@cs.rwth-aachen.de}
}

\maketitle

\begin{abstract}
The increasing digitalization of power grids and especially the shift towards IP-based communication drastically increase the susceptibility to cyberattacks, potentially leading to blackouts and physical damage.
Understanding the involved risks, the interplay of communication and physical assets, and the effects of cyberattacks are paramount for the uninterrupted operation of this critical infrastructure.
However, as the impact of cyberattacks cannot be researched in real-world power grids, current efforts tend to focus on analyzing isolated aspects at small scales, often covering only either physical or communication assets.
To fill this gap, we present \NAME{}, a comprehensive research environment that facilitates reproducing, implementing, and analyzing cyberattacks against power grids and, in particular, their impact on both communication and physical processes. %
We validate \NAME{}'s accuracy against a physical testbed and show its scalability to realistic power grid sizes.
We then perform authentic cyberattacks, such as Industroyer, within the environment and study their impact on the power grid's energy and communication side.
Besides known vulnerabilities, our results reveal the ripple effects of susceptible communication on complex cyber-physical processes and thus lay the foundation for effective countermeasures.

\end{abstract}

\section{Introduction}%
\label{sec:intro}
\acresetall{}

Over the past decades, the communication infrastructure of power grids has shifted from serial communication to Internet-compatible IP-based communication~\cite{mai2020uncharted}.
While this shift provides many benefits, such as enhanced flexibility, adaptability, and scalability~\cite{kwok-hong2002migrating}, it also drastically increases the susceptibility to cyberattacks~\cite{krause2021cybersecurity}.
Past incidents, such as the cyberattacks on the Ukrainian power grid~\cite{whitehead2017ukraine}, %
highlight the severe real-world implications. %
However, besides rather abstract information on these incidents~\cite{whitehead2017ukraine}, little is known about the precise technical susceptibility of power grids and their communication infrastructure w.r.t.\ cyberattacks.
For example, it is unknown which technically feasible attack on the communication infrastructure causes the most drastic impact on the power side and how severe this impact is.
Consequently, academia and industry struggle to identify and prioritize the parts of the power grid communication infrastructure needing countermeasures %
to thwart attacks~\cite{huang2018survey}.

Thus, it is essential to evaluate and understand the impact of cyberattacks on power grids.
Unlike traditional (office or server) networks, the power grid as a cyber-physical system also demands to include the physical, i.e., energy, side in this analysis~\cite{antonioli_minicps_2015}.
However, evaluating cybersecurity in \emph{existing} power grids is typically not feasible due to high availability requirements and the risk of permanently damaging critical physical assets.
Thus, various streams of research study the impact of cyberattacks on \emph{physical} grid operation using testbeds and simulations, e.g.,~\cite{hahn2010development,liu2015integrated,wermann2016astoria}.
From a different angle, related work concerning the security of the \emph{communication} infrastructure of power grids ranges from game-theoretic analyses~\cite{chen2012smart} over \acp{des}~\cite{almajali2012analyzing} to network emulations~\cite{ghosh2017security}.
Still, related work typically does not capture both the communication \emph{and} the energy side in sufficient detail and scale required to thoroughly assess the impact of real cyberattacks on power grids.

To alleviate this situation, other domains report huge success with sophisticated simulation environments, accurately reflecting both the communication and physical side, e.g., for water distribution~\cite{murillo2020cosimulating} and treatment~\cite{antonioli_minicps_2015}, or the Internet of Things~\cite{siboni2019security}.
While these approaches show the feasibility of capturing the combination of communication and the corresponding physical side in sufficient detail, they work in significantly smaller scenarios than typically found in power grids.
Respective power grid approaches often focus on aspects other than cybersecurity, most prominently grid operation (e.g.,~\cite{godfrey2010modeling}).
In turn, when considering cyberattacks, they abstract from actually used communication protocols (e.g.,~\cite{pan2017cyber}) or study only small-scale scenarios, such as individual substations (e.g.,~\cite{tebekaemi2016designing}).
Thus, an environment for the large-scale evaluation of cyberattacks against power grids covering both the communication and energy side has been identified as an important open research field~\cite{henze_cybersecurity_2020,mihal2022smart}.

This paper addresses the challenge of accurately studying the impact of cyberattacks on power distribution grids of realistic size (e.g., 500 electrical assets and 150 communicating parties~\cite{meinecke2020simbench}).
Therefore, we propose \NAME{}, an open-source\footnotemark[1] research environment that combines a state-of-the-art power flow solver with a sophisticated network emulator to flexibly and accurately model the energy and communication side of real-world power grids.

We use \NAME{} to deploy real network applications (e.g., attack tools) within a simulated power grid to comprehensively study the cyber \emph{and} physical impact of attacks.
Our study shows that such cyberattacks can have devastating impacts, particularly when attackers obfuscate immediate effects by manipulating measurements.
Above all, they highlight that complex cyber-physical systems inseparably unite communication and physical processes and thus require a thorough view to analyze attacks.

\noindent
\textbf{Contributions.}
In summary, our contributions are:
\begin{itemize}[noitemsep,topsep=0pt,leftmargin=9pt]

\item We classify research on cyberattacks against power grids, finding that current work tends to focus on the energy \emph{or} communication side, leaving the actual susceptibility of power grids open.
To understand the root cause of this issue, we analyze the state of research environments for power grids and conclude that a comprehensive environment for studying the combined impact of cyberattacks on the energy \emph{and} communication side of power grids at realistic scales is missing (\S\ref{sec:security}).

\item To comprehensively analyze the cybersecurity of power grids, we present \NAME{}\footnote{\url{https://github.com/fkie-cad/wattson} and \url{https://wattson.it}}, an open-source research environment for studying cyberattacks on power grids at realistic scales, allowing us to analyze their impacts on the power \emph{and} networking side through co-simulation.
We validate its accuracy by replicating an attack on a physical testbed and show its scalability to real-world grid sizes using benchmarking and reference grids (\S\ref{sec:design}).

\item We study the impact of authentic, real-world cyberattacks within a reference grid~\cite{meinecke2020simbench} (475 energy and 239 communication assets) within \NAME{}.
Among others, we cover the infamous Industroyer~\cite{industroyer2blogpost} as an actually conducted attack and sophisticated false data injection.
We discussed and validated the practical relevance of these attacks with domain experts from a national cybersecurity agency and multiple grid operators.
Our results show the susceptibility of power grids against sophisticated cyberattacks and that only the combined consideration of energy \emph{and} communication allows extensive analyses of their impact, thus laying the foundation to secure them properly (\S\ref{sec:attacks}).

\end{itemize}

\noindent
\textbf{Availability Statement.}
To spur further research on evaluating the impact of cyberattacks against power grids at realistic scales, \NAME{} is available under an open-source license\footnotemark[1]. 
Furthermore, we provide the network traces and the physical grid states for our conducted attacks\footnotemark[1].

\section{Cybersecurity in Power Grids}%
\label{sec:security}

To lay the foundation for our work, we provide a brief background on power grids introducing their structure and operation (\S\ref{sec:security:background}).
We then classify possible cyberattacks against power grids (\S\ref{sec:security:attacks}) based on a literature survey to identify current shortcomings and derive requirements for comprehensive cybersecurity research in power grids (\S\ref{sec:security:research}).
Finally, we analyze how related work covers this cross-domain research area and fulfills the derived requirements for evaluating cybersecurity in power grids %
(\S\ref{sec:security:rw}). %

\subsection{Background on Power Grids}%
\label{sec:security:background}

The main task of a power grid is to transmit electric power from generators to consumers~\cite{nardelli2014powergrid}.
Therefore, it uses alternating currents with different voltage levels, ranging from low to extra-high voltage, following a hierarchic structure for efficiency and minimizing losses.
While \emph{transmission networks} supply larger industrial consumers and typically operate at higher voltage levels, the subordinated \emph{distribution networks} supply smaller industrial and residential consumers and usually operate on lower voltage levels.
Electrical \emph{substations} connect these different networks and \emph{transform} between the voltages as needed.

A major challenge in power grid operation is maintaining the balance between power demand and supply, as large quantities of electrical power cannot be stored efficiently.
Deviations from the nominal frequency (e.g., \SI{50}{\hertz} in Europe; \SI{60}{\hertz} in the US) indicate an imbalance of demand and supply. %
Furthermore, the capacities of \emph{transmission lines} and other physical assets limit the power flow, where overloading induces physical damage.
\emph{Protective relays} typically prevent such damage, whereas falsely tripping such relays may lead to (partial) blackouts.

Power grids usually rely on SCADA systems to monitor and control the correct operation and adapt to the growing number of digital assets, e.g., enabling \emph{\acp{se}} and remote tripping of relays~\cite{thomas2017power}.
Therefore, distributed \acp{rtu} logically connect %
to a centralized \ac{mtu} for reporting measurements and receiving commands.
Such systems thus depend on a communication infrastructure and suited protocols. %
Due to the increasing number of renewable power sources, they exhibit growing importance.

However, most protocols used to manage power grids, such as IEC 60870-5-104 (in Europe) or DNP3 (in the US), were not designed with security in mind~\cite{krause2021cybersecurity}.
Ongoing digitalization and insecure, hard-to-replace protocols make power grids vulnerable to various cyberattacks~\cite{case2016analysis, james_improving_2019}.
While \ac{ict} security mainly relies on dedicated networks with access restrictions, these networks provide insufficient internal security.
Thus, they are especially vulnerable once attackers gain initial access, e.g., using credential theft, spear phishing, or physical intrusion into substations~\cite{hong2014integrated}. %

The cyberattacks on the Ukrainian power grid in 2015, 2016, and 2022~\cite{case2016analysis,mcfail2022detection,industroyer2blogpost} demonstrated the vulnerabilities of power grids.
In particular, they spur intensive research on understanding individual incidents with devastating consequences for the power grid and ultimately improving countermeasures.
While this paper mainly focuses on the structure and operation of current and future European (\ac{ac}) power grids, its contributions generally also apply to other power grids, such as North American ones.

\subsection{Classification and Analysis of Attacks}
\label{sec:security:attacks}

For structuring and understanding the vast range of cyber threats against power grids, we propose a classification according to the required attackers' knowledge for successfully performing the attack. %
This approach classifies attacks based on their complexity and already supports the consideration of adequate countermeasures.
As a result, we distinguish between \emph{physical}, \emph{syntactic}, and \emph{semantic} attacks, which we introduce in the following while also discussing their assumptions and requirements.

\noindent\textbf{Physical Attacks.}
Physical attacks comprise all incidents where attackers disturb the power grid by physically interrupting, tampering with, or destroying grid equipment, including any \ac{ict} components.
Such attacks range from simple \emph{device disconnects} (e.g., \cite{soltan2017power}) to more complex \emph{demand manipulation}, where attackers simultaneously start up many high-wattage devices (e.g., \cite{huang2019not}).
As a prerequisite, attackers need (physical) access to the targeted devices, while limited knowledge about the power grid's \ac{ict} and processes typically suffices to execute such attacks.

\noindent\textbf{Syntactic Attacks.}
In syntactic attacks, attackers use crafted, intercepted, or duplicated messages to interfere with the power grid's operation \emph{through} the \ac{ict}.
Although these messages are syntactically correct, attackers typically do not need profound process knowledge since they do not modify the messages' payload.
Such attacks often manifest as \emph{\ac{dos}} attacks, targeting the availability of (parts of) a power grid (e.g., \cite{almajali2012analyzing}).
Alternatively, attackers may \emph{replay} messages to provoke unwanted or harmful behavior (e.g., \cite{zhao2016detection}).
Typically, attackers require (remote) access to the \ac{ict} to perform such attacks successfully.

\noindent\textbf{Semantic Attacks.}
Semantic attacks have the same requirements as syntactic attacks, with the addition that attackers require some knowledge about the process, configuration, and topology to target power grid operations specifically.
Then, attackers can unsuspiciously counterfeit measurements or commands within \ac{ict} messages, which may have devastating effects.
Depending on the targeted process, it may even suffice to \emph{replay} specific messages at the right time (e.g., \cite{pavithra2021prevention}).
However, the most prominent attack is \emph{false data injection}, where attackers tamper with measurements to mislead grid operation (e.g., \cite{rawat2015detection}).

\begin{table}
\centering
\caption{Classification of power grid attacks into \underline{Phys}ical, \underline{Syn}tactic, and \underline{Sem}antic and how current research evaluates attack impact w.r.t. ICT and power grid.}
\renewcommand{\arraystretch}{.7}
\begin{tabularx}{\linewidth}{lXm{0.23\linewidth}m{0.31\linewidth}}
\toprule
 & \textbf{Attack Type} & ~\textbf{ICT} & ~\textbf{Power Grid} \\
\midrule
\multirow{4}[2]{*}{\rotatebox[origin=c]{90}{\textbf{Phys.}}} & Device & & \multirow{2}[0]{*}{\cite{soltan2017power,hossain2019line}} \\
& Disconnect & & \\
\cmidrule(l){2-4}
 & Demand & & \cite{soltan2018blackiot,huang2019not} \\
& Manipulation & & \cite{simonov2019detecting,wang2019securing} \\
\midrule
\multirow{3}[2]{*}{\rotatebox[origin=c]{90}{\textbf{Syn.}}} & \multirow{2}[1]{*}{Denial-of-Service} & \cite{mallouhi2011testbed,chen2012smart,almajali2012analyzing} & \textbf{\cite{srikantha2015denial}}, \cite{albarakati2018openstack,hannon_dssnet_2016} \\[1pt]
& & \cite{zaidi2012stochastic}, \textbf{\cite{srikantha2015denial}} & \cite{hasnat2019data,zemanek2022powerduck,lin2012sec-geco} \\
\cmidrule(l){2-4}
 & \multirow{2}[2]{*}{Replay} & \cite{wu2011fault,li2014efficient,liu2016lightweight} & \cite{zhao2016detection,irita2017detection,zemanek2022powerduck} \\
\cmidrule{1-1}
\cmidrule{3-4}
\multirow{4}[2]{*}{\rotatebox[origin=c]{90}{\textbf{Sem.}}} & & \cite{pavithra2021prevention} & \cite{albarakati2018openstack,tran2013detection,irita2017detection} \\
\cmidrule(l){2-4}
& \multirow{3}[1]{*}{False Data Injection} & \multirow{2}*{\cite{kim2011strategic,chen2012smart,kim2013on}} & \cite{albarakati2018openstack,pan2017cyber,lin2012sec-geco,duan2020cybersecurity} \\[1pt]

& & \multirow{2}*{\cite{li2014efficient,kim2014data,wang2020semantic}} & \cite{liu2011false,kosut2011malicious,zhang2013synchronization,dagoumas2019assessing} \\[1pt]
& & & \cite{rawat2015detection,liang2017review,jin2019power,gao2021assessment} \\
\bottomrule
\end{tabularx}
\label{tab:attack-classification}
\end{table}

To understand to which extent current research covers these attack types, we performed a survey of $36$ scientific publications which analyze the impact of attacks against power grids.
We follow the discussed attack classification, i.e., \emph{physical}, \emph{syntactic}, or \emph{semantic} attacks, for comparing and structuring related work in Table~\ref{tab:attack-classification}.
For each publication, we identify the considered impact of the investigated attacks, i.e., \emph{\ac{ict}}, \emph{power grid}, or even both.

Our survey reveals that current research primarily focuses on single attack types, with a few exceptions considering two or three different attacks~\cite{albarakati2018openstack,li2014efficient,lin2012sec-geco}.
However, none covers \emph{all} proposed attack categories.
Importantly, they exclusively focus on the \ac{ict} \emph{or} the energy side, abstracting from the respective other domain, although successful attacks inevitably concern both domains.
As an exception,~\cite{srikantha2015denial} considers both the \ac{ict} and the energy side.
However, with its focus on a \ac{dos} attack, it merely covers semantic attacks.
As its mathematical modeling restricts the approach's potential for analyzing different attack types and relies on assumption-based attack effects, it is inappropriate for studying, e.g., semantic attacks and unforeseeable inter-domain side effects.
Consequently, we argue that a meaningful analysis of cyberattacks on power grids requires a more comprehensive evaluation methodology, adequately reflecting the complex cyber-physical dependencies of the \ac{ict} and energy side, while also covering the entire range of possible attacks.
In the following, we thus derive the requirements for such a methodology before continuing our related work analysis.

\subsection{Requirements for Research Environments}%
\label{sec:security:research}

For adequately studying sophisticated cyberattacks on power grids, an appropriate research environment needs to fulfill specific requirements.
We derive a total of four requirements based on the results of our literature survey in \S\ref{sec:security:attacks} and explain them in the following:%

\noindent\textbf{Accuracy.}
Both the \ac{ict} and power grid components behave and interact according to their respective real components.
This includes \emph{real} communication behavior and the cross-domain interplay between \ac{ict} and power grid components during normal operations and cyberattacks.
As some types of communication in real power grid networks affect the state and behavior of the grid component and, e.g., physical defects of these components can impact the communication behavior, this \emph{cross-domain accuracy} is critical to assess the impact of cyberattacks.

\noindent\textbf{Scalability.}
Research environments provide scalability w.r.t. the realizable scenarios.
This ranges from small laboratory settings to realistic power grids with high complexity, i.e., with significantly more components, for comprehensively evaluating the impact of cyberattacks.

\noindent\textbf{Flexibility.}
\ac{ict} networks, power grid topologies, and cyberattacks are flexible in evaluating various configurations and scenarios.
Further, communication protocols, attack tools, defensive measures, and power grid processes are exchangeable, enabling extensive research opportunities.

\noindent\textbf{Cybersecurity.}
Finally, a suitable research environment explicitly considers cybersecurity analyses in its design.
Through the inclusion of real-world attack tools along with respective evaluation capabilities, studying realistic cyber threats becomes feasible.

\begin{table*}
\renewcommand{\tabularxcolumn}[1]{m{#1}}
\newcolumntype{Y}{>{\centering\hsize=\dimexpr2\hsize+2\tabcolsep+\arrayrulewidth\relax}X}
\renewcommand{\arraystretch}{.6}
\newcolumntype{Z}{>{\hsize=.5\hsize}X}
\centering
\caption{Classification of co-simulation approaches regarding their \underline{com}munication model (continuous or discrete) and their \underline{power} model (steady or transient state) along with their fulfillment of required properties.}
\vspace{-1.1pt}
\begin{tabularx}{\linewidth}{lllccccccccc}
\toprule
\textbf{Com.} & \textbf{Power} & \multirow{2}{*}{\textbf{Approaches}} & \multicolumn{2}{c}{\textbf{Accuracy}} & \multicolumn{2}{c}{\textbf{Scalability}} & \multicolumn{2}{c}{\textbf{Flexibility}} &
\multicolumn{2}{c}{\textbf{Cybersecurity}} & \textbf{Open}\\
\cmidrule(lr){4-5}
\cmidrule(lr){6-7}
\cmidrule(lr){8-9}
\cmidrule(lr){10-11}
\textbf{Model} & \textbf{Model} & & \textbf{Com.} & \textbf{Power} & \textbf{Com.} & \textbf{Power} & \textbf{Com.} & \textbf{Power} & \textbf{Com.} & \textbf{Power} & \textbf{Source} \\
\midrule

\multirow{8}[14]{*}{\textbf{Discrete}}
& \multirow{4}[6]{*}{\textbf{Steady}}
& \cite{bhor2016network}, \cite{palmintier2017helics} & \rate{2} & \rate{4} & \rate{5} & \rate{5} & \rate{3} & \rate{4} & \rate{1} & \rate{1} & \rateYes{}
\\ \cmidrule(l){3-12}
& & \cite{duan2020cybersecurity}, \cite{souza2020cosim} & \rate{2} & \rateU{3} & \rate{5} & \rateU{5} & \rate{3} & \rate{4} & \rate{2} & \rate{4} & \rateNo{}
\\ \cmidrule(l){3-12}
& & \cite{mallouhi2011testbed} & \rate{2} & \rateU{3} & \rate{5} & \rateU{5} & \rate{3} & \rate{4} & \rate{4} & \rate{1} & \rateNo{}
\\ \cmidrule(l){3-12}
& & \cite{celli2014dms}, \cite{garau2017smart}, \cite{godfrey2010modeling}, \cite{levesque2012sim}, \cite{mets2011integrated} & \rate{2} & \rateU{3} & \rate{5} & \rateU{5} & \rate{3} & \rate{4} & \rate{1} & \rate{1} & \rateYes{}
\\ \cmidrule(l){2-12}

& \multirow{4}[5]{*}{\textbf{Transient}}
& \cite{amarasekara2015cosim}, \cite{bian2015realtime}, \cite{georg2014inspire}, \cite{hopkinson2006epochs}, \cite{li2011vpnet}, & \multirow{2}[1]{*}{\rate{2}} & \multirow{2}[1]{*}{\rateU{?}} & \multirow{2}[1]{*}{\rate{5}} & \multirow{2}[1]{*}{\rateU{4}} & \multirow{2}[1]{*}{\rate{3}} & \multirow{2}[1]{*}{\rate{4}} & \multirow{2}[1]{*}{\rate{1}} & \multirow{2}[1]{*}{\rate{1}} & \multirow{2}[1]{*}{\rateNo{}}
\\[2pt]
& & \cite{liberatore2011smart}, \cite{shum2018cosim}, \cite{veith2020largescale} & & & & & & &
\\ \cmidrule(l){3-12}
& & \cite{ciraci2014fncs}, \cite{hansen2017eval}, \cite{kelley2015federated}, \cite{nutaro2007integrated,nutaro2011power} & \rate{2} & \rateU{?} & \rate{5} & \rateU{5} & \rate{3} & \rate{4} & \rate{1} & \rate{1} & \rateYes{}
\\
\cmidrule(l){3-12}
& & \cite{lin2012geco,lin2012sec-geco}, \cite{pan2017cyber} & \rate{2} & \rateU{3} & \rate{5} & \rate{5} & \rate{3} & \rate{4} & \rate{1} & \rate{5} & \rateNo{}
\\
\midrule
\multirow{4}[3]{*}{\textbf{Continuous}}
& \multirow{2}[2]{*}{\textbf{Steady}}
& \cite{hannon_dssnet_2016,hannon2018combining} & \rate{5} & \rate{4} & \rate{3} & \rateU{4} & \rate{4} & \rateU{4} & \rateU{2} & \rate{5} & \rateYes{}
\\ \cmidrule(l){3-12}
& & \cite{li2017distributed} & \rate{5} & \rate{4} & \rate{4} & \rate{5} & \rateU{4} & \rateU{4} & \rate{1} & \rate{1} & \rateYes{} \\

\cmidrule(l){2-12}
& \textbf{Transient}
& \cite{albarakati2018openstack} & \rateU{4} & \rate{5} & \rateU{3} & \rateU{4} & \rateU{4} & \rateU{4} & \rate{4} & \rate{5} & \rateNo{}
\\
\specialrule{1pt}{.2pc}{.5pc}
\textbf{Continuous} & \textbf{Steady} & \textbf{\NAME{}} & \rate{5} & \rate{4} & \rate{4} & \rate{5} & \rate{5} & \rate{5} & \rate{5} & \rate{5} & \rateYes{} \\

\specialrule{0.5pt}{.5pc}{0pt}
\end{tabularx}
\label{tab:rw-analysis}
\vspace{0.2em}
\newline
\footnotesize Requirement not \rate{1}\,, marginally \rate{3}\,, mostly \rate{4}\,, or thoroughly \rate{5} fulfilled
\hfill\small $*$ -- \footnotesize Not evaluated by authors / uncertain
\hfill ? -- Unknown
\end{table*}

Although individually achieving these design properties is already challenging, a particular challenge arises when fulfilling all of them as required.
We thus continue with our related work discussion, analyzing to which extent current research environments for power grids cover the requirements with the intended depth.

\subsection{Related Work on Research Environments}
\label{sec:security:rw}

Evaluating cyberattacks and countermeasures is typically not possible in actually deployed power grids due to the high availability requirements of such systems and the risk of causing physical defects~\cite{sun2018cyber}.
Although physical testbeds, i.e., deployments of real power grid components for test and development purposes, are optimal regarding the provided accuracy, they are limited in scalability and flexibility~\cite{cintuglu2017survey}.
Consequently, research largely depends on modeling power grid components to analyze and evaluate cybersecurity.
In turn, analytical models of power grids, such as~\cite{liu2013framework,shuvro2017modeling}, offer a scalable and flexible way to analyze the impact of cyberattacks while lacking the required accuracy due to their abstract models.
Hence, a better-suited approach is to simulate or emulate power grid components, enabling a comprehensive representation of the power grid.
Here, a class of approaches exclusively focuses on simulating either the \ac{ict}, such as~\cite{liu2015analyzing,tebekaemi2016designing,radoglou2019attacking}, or the power grid, such as~\cite{poudel2017real,tolbert2020reconfigurable}.
However, such approaches are extremely limited for cybersecurity research, as discussed in \S\ref{sec:security:attacks}.

Thus, numerous approaches aim to \emph{co-simulate} the \ac{ict} and the power grid.
Such approaches follow general \ac{ict} and power simulation paradigms:
A communication network simulation either uses a \emph{discrete event} or a \emph{continuous} approach.
While \acfp{des} allow non-real-time simulations and offer high scalability, they cannot precisely represent actual network traffic, involved protocol stacks, and timing-related side effects due to the chosen level of abstraction~\cite{zhang2021smart}.
As a result, they have limited accuracy and flexibility w.r.t. the evaluation of cyberattacks.
Here, continuous approaches are favorable while having potential drawbacks regarding performance.

Similarly, for power grid simulations, a distinction is made between \emph{steady state} and \emph{transient state}.
While the former only considers the final, i.e., steady, state after the system's parameters change, transient state simulation also models the process, i.e., the transient behavior, between these steady states.
Transient state simulation thus offers more insights at the cost of higher complexity for model parameters and calculations~\cite{balduin2019towards}.
In Table~\ref{tab:rw-analysis}, we classify different co-simulation approaches based on these paradigms and summarize to which extent they fulfill the requirements derived in \S\ref{sec:security:research}.
We further indicate their (desirable) open-source availability, i.e., that they do not rely on proprietary soft- or hardware.

Most approaches rely on discrete-event simulation (DES), offering scalable means for analyzing large networks.
However, discrete-event simulation requires precise prior knowledge about communication behavior and potential side effects~\cite{wehrle2010modeling}.
This is particularly challenging for cybersecurity research, where attacks occasionally attempt to undermine such assumptions and where the communication behavior is often part of the research subject.
Further, discrete models for all involved tools and applications are required, which is difficult when including real programs, e.g., malware.
Hence, we assess that \ac{des} lacks the required accuracy and flexibility for conducting sophisticated cybersecurity research in power grids, i.e., covering the entire scope of possible attacks (cf.~\S\ref{sec:security:attacks}).

On the power grid side, a transient model offers a more precise representation of the grid's behavior while considering time-based effects.
Such transient effects are particularly interesting when modeling safety equipment that protects power grid components from excessive currents or voltages.
However, transient simulations require precise knowledge about the components' characteristics while being computationally expensive, thus often requiring dedicated hardware.
Furthermore, the gained detail level can be neglected in many cases~\cite{balduin2019towards}.
Therefore, steady-state simulations are similarly well-suited for evaluating the power grid's state and behavior while being computationally less expensive.

\begin{figure*}[t]
    \centering
    \includegraphics[width=\linewidth]{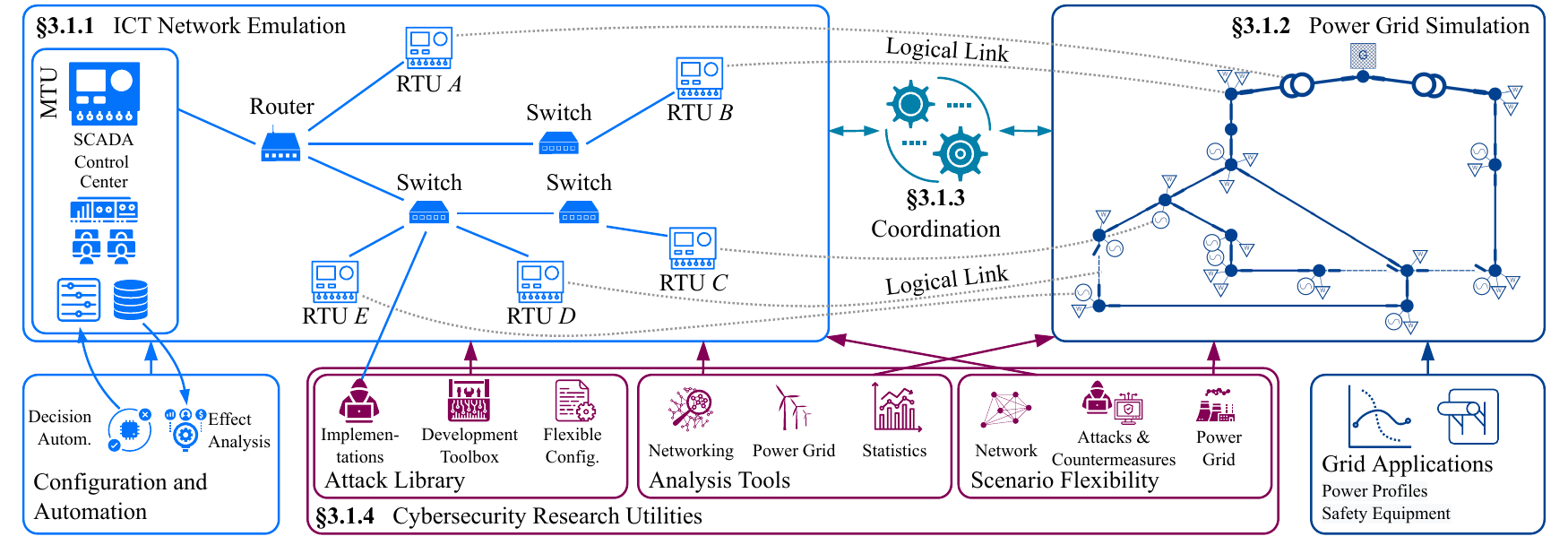}
    \caption{\NAME{}'s architecture comprises three main components: an \ac{ict} network emulation, a power grid simulation, and a coordinator, complemented by several cybersecurity research utilities.
    The \ac{ict} network emulation deploys the network topology, emulates switches, links, routers, and hosts, and further supports the attachment of different applications, e.g., attackers or \acp{ids}.
    \acp{rtu} are linked to the power grid via the coordinator and actively control and monitor changes within the power grid simulation, which can be configured with, e.g., power profiles and safety equipment.
    }
    \label{fig:architecture}
\end{figure*}

Explicitly focusing on approaches that rely on continuous communication network simulation (as favorable for analyzing the impact of cyberattacks),
we find that Albarakati et al.~\cite{albarakati2018openstack} propose a transient approach where the co-simulator is based on OpenStack and OPAL-RT. %
However, the lack of validation of their approach's overall accuracy and scalability and the involvement of specialized and costly hardware exacerbates an independent evaluation of these critical requirements.
Moreover, the proprietary power simulator and the involvement of hardware-based \ac{ict} components limit the flexibility of their approach regarding the range of possible scenarios and cyberattacks.

In turn, DSSnet~\cite{hannon_dssnet_2016,hannon2018combining} follows a continuous, steady-state approach integrating Mininet and OpenDSS, which, however, suffers from poor scalability due to significant coordination overheads.
Although DSSnet includes a continuous communication model, the authors confine themselves to simulating only the potential effects of cyberattacks instead of realistically conducting such attacks, which limits the assessability of the achieved flexibility and aptitude for cybersecurity research.
Similarly, the co-simulator proposed by Li et al.~\cite{li2017distributed} uses CORE (continuous) and GridLabD (steady state) to model \ac{ict} and power grid, respectively.
While their distributed approach offers scalability, the authors neither validate its accuracy nor its suitability for cybersecurity evaluation.

To summarize, multiple co-simulation approaches address the challenge of enabling cross-domain research within power grids, although most do not specifically focus on cybersecurity research.
Thus, design decisions specific to particular use cases, e.g., using a discrete communication model for scalability, impede their applicability for sophisticated cybersecurity research, as they cannot, amongst others, accurately implement real-world attacks.
Further, the observed focus on either the communication or the power side (cf.~\S\ref{sec:security:attacks}) similarly holds for most research environments.
Those few approaches that potentially offer the required cross-domain accuracy involve costly (proprietary) hardware, offer limited scenario flexibility, or exhibit indeterminate scalability.
Consequently, a suitable research environment adhering to all requirements of \S\ref{sec:security:research} is still needed to facilitate comprehensive analyses of cyberattacks on power grids.

\section{\NAME{} -- A Foundation for Analyzing the Impact of Cyberattacks on Power Grids}
\label{sec:design}
Despite the sophisticated approaches for evaluating the impact of cyberattacks on power grids offered by related work, our analysis underlines the lack of cybersecurity-focused environments that accurately and flexibly cover the \ac{ict} and energy side while maintaining scalability to realistic grid sizes.
Hence, we introduce \NAME{}%
, a cybersecurity research environment for power grids striving to achieve these properties (cf.~\S\ref{sec:security:research}) by combining a detailed \ac{ict} network emulation with a flexible power grid simulation and integrated cybersecurity research tools, covering attack frameworks, evaluation tools, and the possibility to deploy realistic countermeasures.
\NAME{} combines valuable concepts from related work, while its cybersecurity-focused architecture makes it unique compared to these approaches.
In the following, we present \NAME{}'s design (\S\ref{sec:design:architecture}) before evaluating its accuracy, scalability, flexibility, and cybersecurity features~(\S\ref{subsec:architecture-performance}).

\subsection{Architecture Design}
\label{sec:design:architecture}

We depict \NAME{}'s architecture and components in Fig.~\ref{fig:architecture}, consisting of a realistic \emph{\ac{ict} network emulation}, an accurate \emph{power grid simulation}, a dedicated \emph{coordinator} linking and synchronizing the two, and \emph{cybersecurity research utilities}.
Hence, \NAME{} enables analyzing the interaction and interplay of the communication network (i.e., hosts, routers, switches, links) and the power grid (e.g., transformers, lines, loads, or generators).
Besides a manual configuration of the \ac{ict} network, \NAME{} supports automatically deriving grid configurations through a dedicated modeling approach~\cite{klaer_modeling_2020}.
\NAME{} is primarily implemented in Python and we refer to our documentation website\footnote{\url{https://wattson.it}} for technical details.
In the following, we detail the design of \NAME{}'s components.

\subsubsection{ICT Network Emulation}
To facilitate \emph{flexible} network topologies and the deployment of universal network-based applications, \NAME{} uses network emulation to model the \ac{ict} network of power grids.
This approach enables the use of realistic communication protocols, e.g., IEC~60870-5-104, network monitoring tools such as Wireshark, and network-based attackers.
We utilize Containernet~\cite{peuster2016medicine}, a fork of the network emulator Mininet~\cite{lantz2010network}, to create an \ac{ict} network comprising switches, routers, links, and process- and Docker-based hosts.
The emulator uses Linux network namespaces to create several virtual hosts on the same physical host along with virtual switches based on Open vSwitch~\cite{pfaff2015design} and virtual links.
Thereby, the precise configuration of devices and the underlying network, including link properties, e.g., jitter, bandwidth and delay, and realistic communication using the Linux networking stack down to Layer~2 become possible.

We extend Containernet to realize horizontal \emph{scalability} of the network emulation by partitioning the network into several segments that can be distributed onto several physical hosts and connected via physical links.
For emulating the \ac{ict} network model, we include implementations of \acp{rtu} and the corresponding \ac{mtu} using the IEC~60870-5-104 protocol, and a graphical \ac{vcc}.
\acp{rtu} serve as \acp{ied} attached to one or multiple grid components.
They transmit monitoring information, e.g., voltage measurements, to the \ac{mtu} at the control center, which, in turn, can issue control commands to monitor and actively manage the grid.
All components realistically communicate over real networking protocols (Ethernet, IPv4, TCP, IEC~104).
Thus, \NAME{} supports the evaluation of realistic \ac{ict} attacks such as \ac{dos}, network reconnaissance, \ac{arp} spoofing, and \ac{mitm} attacks.

\subsubsection{Power Grid Simulation}
\NAME{} uses a steady-state simulation for the power grid, achieving \emph{scalability} and meeting the real-time requirements induced by network emulation.
Here, we utilize pandapower~\cite{pandapower2018}, a power flow solver supporting symmetric \ac{ac} single- and three-phase systems.
Consequently, changes in the configuration of the power grid, e.g., power adjustments, consumption changes, or topology changes resulting from opening or closing circuit breakers, trigger a simulation step that outputs the grid's power flow once a steady state is reached.
Moreover, \NAME{} supports adding Gaussian noise to power values of loads and generators before and after the power flow computation to model slight control deviations and measurement inaccuracies.
Further, it provides load profiles representing the realistic behavior of the power demand over time.

While the steady-state approach allows the power simulation to keep up with the real-time network emulation, \NAME{} cannot directly include any transient compensation processes.
Still, we assess that the scalability and real-time capabilities of this approach offsets this imperfection and that the steady-state simulation suffices for comprehensively researching realistic cyberattacks.

\subsubsection{Simulation Coordination}
\emph{Accurately} representing cyberattacks on power grids requires the consideration of the interplay between networking devices and physical grid components.
Thus, we have to link the network emulation with the power grid simulation and coordinate their respective interactions.
\NAME{} includes a dedicated coordinator, interfacing with each host within the emulated network for interacting with the power grid simulation.
In particular, each host is connected to a dedicated \emph{management network} that solely transmits coordination traffic.
By communicating with the coordinator, hosts can receive the grid's current state and update its configuration.
As soon as the coordinator receives a request to change the grid configuration, it triggers the power simulation.

The functional linking of networking hosts to power grid components (cf. Fig.~\ref{fig:architecture}) is realized by a respective configuration of hosts, either manually or using an automated modeling approach~\cite{klaer_modeling_2020}.
For each monitoring and control information, the configuration defines the responsible \ac{rtu}, the IEC~104 \ac{ioa}, and information for interacting with the coordinator and the respective grid element.
Through its close coupling with the power simulator, the coordinator further enables \NAME{} to apply flexible logic applications to the power grid, e.g., load profiles, self-adjusting inverters, models for safety devices, or defects between simulation steps. %

\setcounter{figure}{2}
\begin{figure*}[b]
    \includegraphics[width=\linewidth]{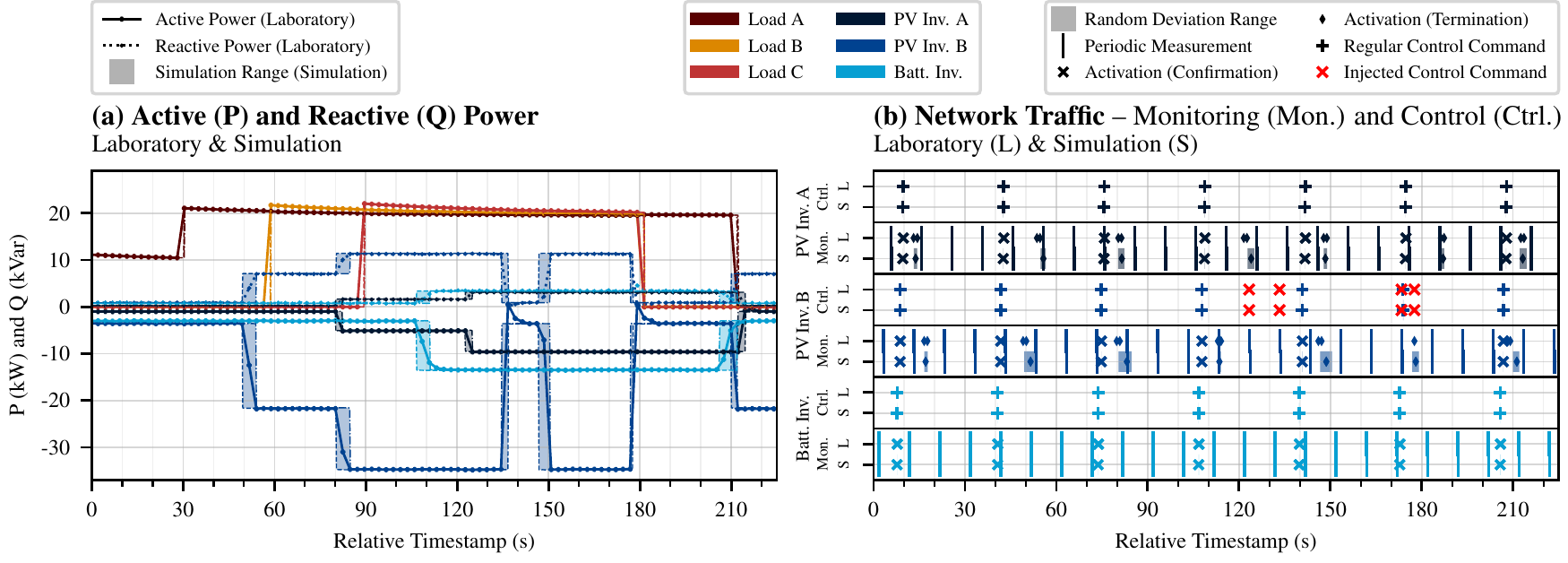}
    \caption{To validate \NAME{}'s accuracy, we compare its behavior to a physical testbed for power grid and \ac{ict} networking under normal and attack conditions.
    For the power grid, the simulation precisely represents the testbed behavior, as course and absolute values of power measurements correspond to real-world measurements.
    Similarly, \NAME{} matches the communication patterns and contents of the testbed network and its devices.}
    \label{fig:accuracy}
\end{figure*}

\subsubsection{Cybersecurity Research Utilities}
We design \NAME{} as a thorough cybersecurity research environment, going beyond the scope of a pure co-simulator.
For every component, we focus on including cybersecurity-related tools and functionality as well as their flexible extendability for future research directions.
Particularly, \NAME{} integrates three design aspects directly contributing to its cybersecurity focus.

\emph{Comprehensive Attack Library.}
\NAME{} comes with an integrated cyberattack framework.
Ranging from various \ac{dos} attack variants over re-implementations of existing malware, e.g., the Industroyer, to a sophisticated library for transparent semantic, e.g., false data injection, attacks, \NAME{} provides a complex toolbox for conducting and evaluating the impact of such attacks in different scenarios.
Due to ethical and risk considerations, we refrain from publishing this attack library.

\emph{Integrated Analysis Tools.}
\NAME{} includes analysis tools per design to provide insight into the process and results of attacks and potential countermeasures.
Based on configurable capture and export tools, e.g., power grid exports and targeted packet captures, \NAME{} provides a library for analyzing these artifacts.

\emph{Scenario Flexibility.}
Finally, \NAME{} targets to enable \emph{flexible} research opportunities.
Hence, we implement extensible scenarios, where changes and extensions to the power grid and the communication network's topologies and behaviors, including attack and countermeasure deployments, can be straightforwardly realized.

To summarize, \NAME{} leverages two specialized tools for power flow computation and network emulation, coupled by a dedicated coordinator and enriched with advanced cybersecurity research tools.
It thus enables the deployment and execution of real networking protocols and applications within the \ac{ict} network that interact with the power grid.
We now analyze and discuss \NAME{}'s fulfillment of the demanded requirements (cf.~\S\ref{sec:security:research}). %

\subsection{Fulfillment of Requirements}
\label{subsec:architecture-performance}

For \NAME{} to be usable and applicable to study the impact of cyberattacks on power grids, we need to ensure it fulfills the derived requirements, i.e., \emph{accuracy}, \emph{scalability}, \emph{flexibility}, and suitability for \emph{cybersecurity} research.
In the following, we thus validate \NAME{}'s accuracy by replicating an attack on a physical testbed.
Then, we show that \NAME{} scales to realistic grid sizes based on benchmarking scenarios and reference grids.
Finally, we discuss its flexibility and cybersecurity aspects.

\subsubsection{Accuracy}
To analyze \NAME{}'s accuracy, we replicate a physical low voltage distribution grid testbed operated at RWTH Aachen University~\cite{velde2020medit,sen2021investigating} shown in Fig.~\ref{fig:laboratory-setup}
within \NAME{} and compare the behaviors of simulation and testbed. 
Hereby, we rely on the correct operation of both, Containernet~\cite{peuster2016medicine} and pandapower~\cite{pandapower2018}, analyzing \NAME{}'s combined, i.e., overall accuracy.
The testbed comprises a medium/low voltage substation at \SI{630}{\kilo\voltampere}, two \ac{pv} inverters (Inv.), a battery inverter, three resistive loads at \SI{20}{\kilo\watt} maximum power consumption, and a corresponding \ac{ict} network.
While all three inverters are controlled via IEC~104, the power consumption of the loads follows a specified time series power profile and is not controlled by the grid operator.
During the experiment, the power output of the inverters is adjusted to match the respective demand. %
All inverters maintain a minimum power output %
and apply a predefined power factor of \num{0.95}.
The \ac{ict} network comprises a single \ac{mtu}, three \acp{rtu}, three switches, and an attack host attached to the switch at the \ac{rtu} of the \emph{PV Inv. B}.

At the beginning of the experiment, the attackers perform an \ac{arp} spoofing attack against said \ac{rtu} and the \ac{mtu} to redirect the traffic between these hosts.
At \SI{123}{\second} and \SI{173}{\second} into the experiment, the attackers inject a control command setting the power output of the \emph{PV Inv. B} to its minimum value. %
There is a non-neglectable delay between the reception of the  %
command and the visible realization in the measurements for all three inverters in the laboratory setup.
We observe a random delay in the physical \acp{rtu}, resulting in a slight uncertainty regarding the actual delay due to measurement intervals of \SI{\approx 2}{\second}.
Thus, we determine the earliest and latest command realization time based on the power measurements and model this uncertainty in the simulation by applying a random delay between the minimum and maximum real delay. %
We conduct \num{20} simulation runs of the physical testbed scenario, each using a different genuine random seed~\cite{hotbits}.
We plot the laboratory and simulation measurements for active and reactive power and network communication in Fig.~\ref{fig:accuracy}.
For the simulation, the plot indicates the minimum and maximum measurement and highlights the area of uncertainty.

\setcounter{figure}{1}
\begin{figure}
    \centering
    \includegraphics[width=\linewidth]{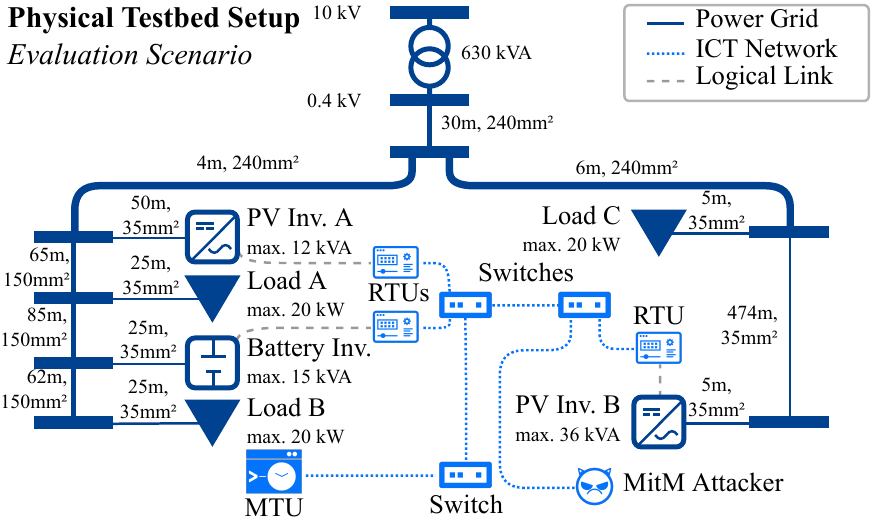}
    \caption{The physical testbed used to validate \NAME{}'s correctness contains a single transformer, two inverters, one battery inverter, and three resistive loads.
    Each inverter can be controlled by a single \ac{mtu} via a dedicated \ac{rtu} from the \ac{ict} network.
    An attack host is attached to one of the switches to perform an \ac{arp}-spoof-based \ac{mitm} attack.
    }
    \label{fig:laboratory-setup}
\end{figure}

\setcounter{figure}{4}
\begin{figure*}[b]
    \includegraphics[width=\linewidth]{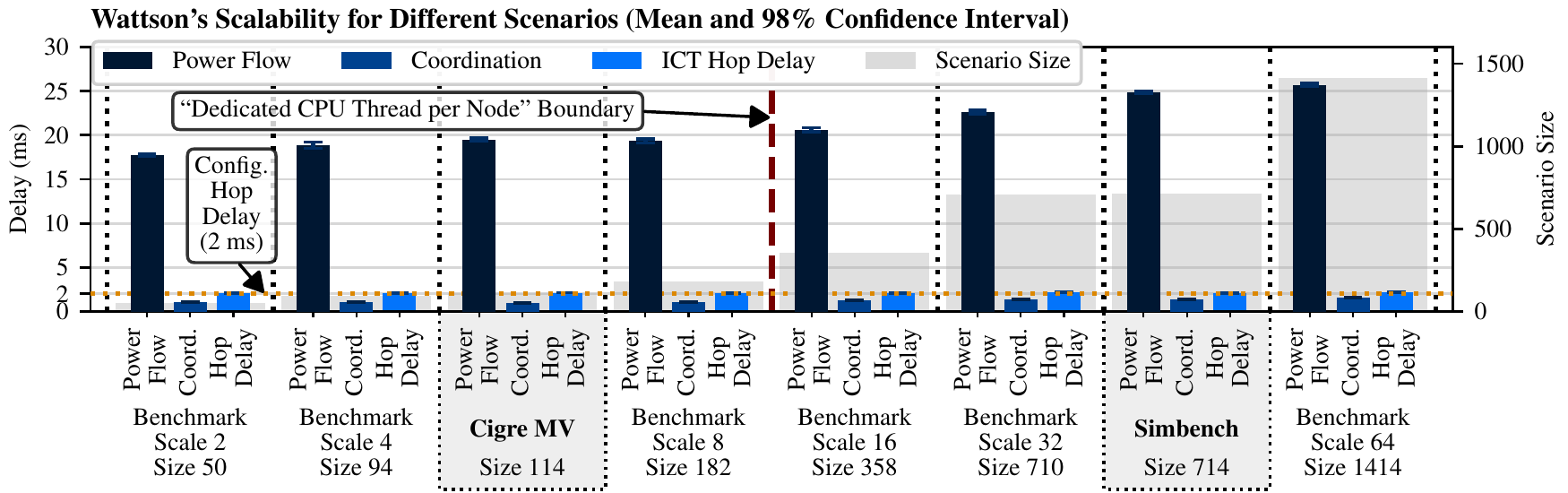}
    \caption{Despite increasing scenario sizes (growing number of assets in the power grid and nodes in the \ac{ict} network), our measurements indicate \NAME{}'s ability to scale to realistic grid size.
    We observe slightly increasing overheads for power flow computation, a reasonably low coordination overhead,
    and \ac{ict} network hop delays corresponding to the actual hop delay of \SI{2}{\milli\second}, enabling us to evaluate cyberattacks at realistic scales.
    }
    \label{fig:performance}
\end{figure*}

\setcounter{figure}{3}
\begin{figure}
    \centering
    \includegraphics[width=\linewidth]{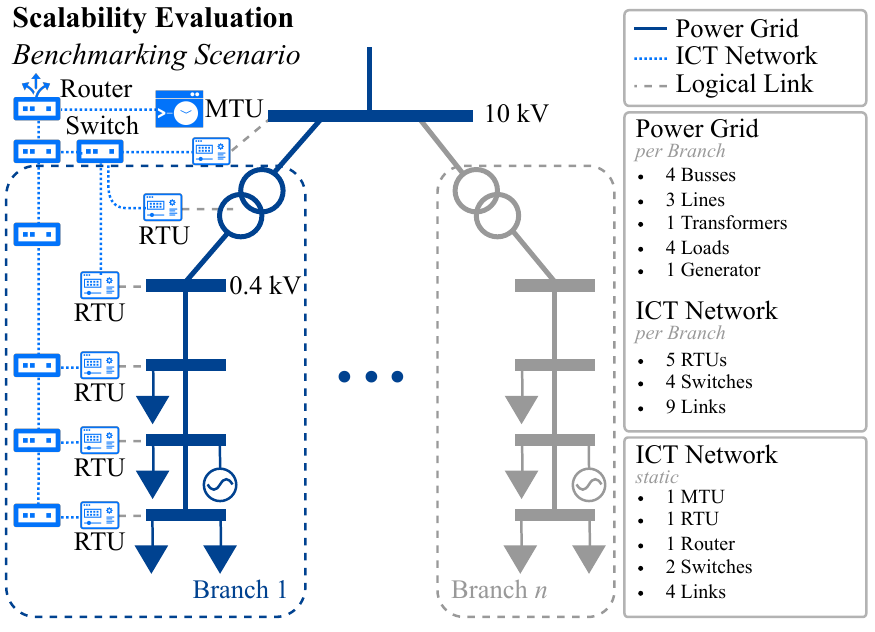}
    \caption{To evaluate \NAME{}'s scalability, we create linearly scalable power grids with $n$ branches, each containing a single transformer, multiple buses, multiple assets, and corresponding \ac{ict} network components, resulting in a linearly growing number of hosts, switches, and links.
    }
    \label{fig:eval-scenario}
\end{figure}
\setcounter{figure}{5}

As shown in Fig.~\ref{fig:accuracy}(a), the simulation corresponds to the laboratory for both active and reactive power.
We observe that the simulation falls within the uncertainty area over multiple runs during power level switching.
Besides these time ranges, the simulation's active and reactive power measurements, voltages, and currents accurately match those from the laboratory.
Notably, these consistencies are not limited to normal operations but continue during the \ac{ict}-driven attack.
For the power level adjustments of the \emph{PV Inv. B} induced by maliciously injected control commands, we observe dips of \SI{\approx4}{\kilo\watt} in the laboratory measurements, which are not present in the simulation.
These %
effects result either from overreactions of the inverter or minor measurement inaccuracies and are not observable with a steady-state simulator.
Apart from these minor deviations, the behavior of the simulated power grid fully corresponds to the laboratory setup.

Similarly, we observe matching traffic patterns between the emulated and the laboratory \ac{ict} network in Fig.~\ref{fig:accuracy}(b), where we indicate timing variations between individual runs with faded areas.
For all three \acp{rtu}, the periodically sent control commands are timed consistently.
Analogously, the monitoring direction is equally present in every periodic measurement in the simulated network.
Except for the variations related to the randomized PLC delays, the deviation areas exhibit a median of only \SI{56}{\milli\second}, indicating \NAME{}'s suitability for
accurately modeling power grids and their corresponding \ac{ict} networks.

\subsubsection{Scalability}
We evaluate \NAME{}'s scalability to show that it can simulate realistic, large-scale power grids, covering architecture-induced overheads and corresponding delays.
For our measurements, we use a single machine with two AMD EPYC 7551 processors, each offering \num{32} cores with \num{2} threads per core and \SI{256}{\gibi\byte} of RAM.
Besides widely-used realistic reference high/medium voltage grids (Cigre MV~\cite{strunz2014benchmark} and Simbench \texttt{1-MV-semiurb--0-sw}~\cite{meinecke2020simbench}), we design specific benchmarking scenarios to show the scalability of \NAME{}.
As shown in Fig.~\ref{fig:eval-scenario}, we use power grid layouts with \(n\) branches (\(n = 2, ..., 64 \)), all connected to a single medium voltage bus.
Each branch contains four low-voltage buses with a linked \ac{rtu} each, a transformer, four loads, a generator, and an \ac{rtu} for the transformer, resulting in a linearly growing number of hosts, switches, and links.
We express the combined number of assets within the \ac{ict} network and power grid as scenario size $S$.

We then investigate
(i) the IEC~104 \emph{communication latency} induced by network stacks and link delays;
(ii) the \emph{coordination overhead} from the delay between \acp{rtu} and coordinator; and
(iii) the \emph{power flow computation} time needed to simulate the grid state.
Here, the \emph{communication latency} per hop should closely adhere to the defined link delays of \SI{2}{\milli\second} plus a minor overhead induced by network stacks. %
The \emph{coordination overhead} should be as low as possible and independent of scenario sizes.
Finally, the \emph{power flow computation} must not exceed delays induced by physical components in real-world power grids, ranging from tens of milliseconds up to \SI{10}{\second}~\cite{base2013load}.

To obtain reliable measurement results, we conduct \num{10} independent runs for each power grid layout (Cigre MV, Simbench \texttt{1-MV-semiurb--0-sw}, and our benchmarking grids with \(2, 4, ..., 64 \) branches).
During each run, the \ac{mtu} performs \num{10} rounds of read and control commands, where all data points referring to bus voltages are read while all generators are set to active power outputs of \SI{50}{\percent} or \SI{100}{\percent}, alternating between rounds.
Hence, the resulting traffic should scale with the size of the power grid while control commands trigger power flow computations.

We show our measurement results in Fig.~\ref{fig:performance}, depicting the arithmetic mean over all runs with the \SI{98}{\percent} mean confidence interval.
First, we ascertain that power flow computations logarithmically scale with increasing grid size, exhibiting adequate scalability of the power simulation.
Similarly, the coordination overhead is reasonably low, with a maximum mean coordination overhead of \SI{1.5}{\milli\second}.
Hence, we assess that the simulation-induced overheads of power flow computations and coordination are suitable to match real-world deployments and thus enable conducting and evaluating cyberattacks at realistic scales.

Moreover, we observe strict adherence of the mean hop delay to the configured link delay of \SI{2}{\milli\second}, with a maximum mean of \SI{2.2}{\milli\second} and a maximum \SI{98}{\percent} confidence interval width of \SI{0.12}{\milli\second} across all scenarios.
Here, the delay slightly increases with growing scenario sizes due to scheduling effects when more nodes need to be simulated than CPU threads are available. %
These numbers are more than sufficient to evaluate cyberattacks at realistic scales.
Still, for larger scenarios or less powerful hardware, \NAME{} supports horizontal scalability by distributing the network over multiple systems.

\subsubsection{Flexibility \& Cybersecurity}
\label{subsec:flexibility}
To tap \NAME{}'s full potential, provided by its accuracy and scalability, we must ensure that its cybersecurity research capabilities offer extensive flexibility.
Here, \NAME{} addresses flexibility requirements on various levels.
First, neither the \ac{ict} nor the power grid topology are fixed or restricted.
They can be freely defined independently of each other, laying the foundation for flexible research scenarios.
Second, their individual behavior as well as their coupling can be flexibly configured, e.g., the implementations of \acp{rtu} and any other host are freely exchangeable, and monitoring and control behavior can be adjusted as needed.
In addition to these general flexibility aspects, \NAME{} further offers flexibility regarding the research questions.
Its network emulation approach allows for arbitrary attack implementations, ranging from custom-made attacks to deployment of real-world malware, whereby \NAME{} already includes an exhaustive library for such attacks.
Combined with capabilities for deploying potential countermeasures, e.g., \acp{ids} or safety measures, \NAME{} provides an extensive toolbox for various research scenarios, such that we assess its flexibility as suitable for cybersecurity research in power grids.

To summarize, our evaluation shows that \NAME{} \emph{accurately} replicates experiments in a physical testbed and can perform real-time simulations of reference and benchmarking scenarios with sufficiently low overhead, thus being able to \emph{scale} to realistic power grid sizes.
Further, it enables \emph{flexible} topologies for the \ac{ict} network and power grid to conduct various customized cyberattacks and deploy arbitrary countermeasures.
These combined properties underline \NAME{}'s suitability for sophisticated \emph{cybersecurity} research in power grids.

\section{Analyzing the Impact of Cyberattacks}
\label{sec:attacks}
Our evaluation shows that \NAME{} fulfills the requirements for performing sophisticated cybersecurity research for power grids.
Thus, we implemented several attacks using \NAME{}, where we followed the attack classes from \S\ref{sec:security:attacks} and discussed and validated their practical relevance with domain experts from a national cybersecurity agency and multiple grid operators.

For all attacks, we rely on the realistic Simbench~\cite{meinecke2020simbench} \texttt{1-MV-semiurb--0-sw} medium voltage distribution grid, representing a suburban grid with two transformers connecting a \SI{110}{\kilo\volt} grid to a \SI{20}{\kilo\volt} grid, resulting in a total of \num{475} assets.
The corresponding \ac{ict} network includes \num{119} \acp{rtu}, a total network node count of \num{239}, resulting in a scenario size of \num{714}.
We define a link latency of \SI{2}{\milli\second} and a bandwidth of \SI{100}{\mega\bit/\second}.

\textbf{Threat Model.} We assume that attackers gained access to the \ac{ict} network either by breaking into an insufficiently secured facility, e.g., an unmanned substation, or through a remote attack, e.g., spear-phishing~\cite{krause2021cybersecurity}.
Thus, the attackers cannot fully control \emph{where} they gain access to the network.
However, they might use network reconnaissance to gather information about vulnerable devices and services to plan and execute the attack.

We begin our analysis with physical attacks, i.e., disconnecting or destroying equipment (\S\ref{sec:attacks:physical}).
Then, we conduct \ac{dos} and \ac{arp} flooding attacks to analyze the impact of syntactic attacks (\S\ref{sec:attacks:syntactical}).
Finally, we study semantic attacks where the attackers manipulate communication to disrupt the safe operation of the power grid (\S\ref{sec:attacks:semantic}).

\subsection{Physical Attacks on Grid Equipment}
\label{sec:attacks:physical}

\begin{figure}
    \includegraphics[width=\linewidth]{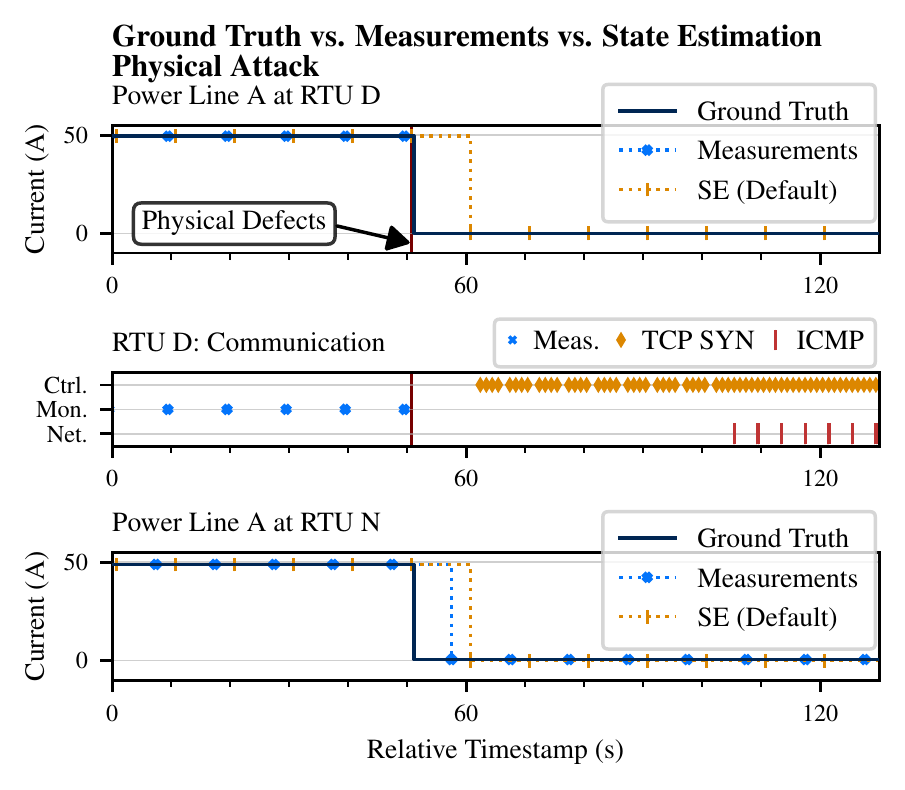}
    \caption{Physical attacks, such as disconnecting a substation (RTU~$D$, transmission lines, and the bus bar), lead to a local blackout (top) and interrupted communication between the substation and the control center (middle). However, the adjacent RTU~$N$ measures the outage and informs the control center (bottom), leading to a correct \ac{se}.}
    \label{fig:physical-attack}
\end{figure}

The cross-domain representation of the power grid within \NAME{} allows for studying the interplay of \ac{ict} and power grid components during an attack.
As an example scenario, we implement a physical attack against a local substation, where the attackers physically disconnect or destroy \ac{ict} devices, i.e., the substation's \ac{rtu} $D$, and local power grid equipment, i.e., transmission lines and the bus bar.
Thus, the substation becomes inoperative for controls issued on-site and remotely.
We plot the resulting effects on the power grid and communication in Fig.~\ref{fig:physical-attack}.

\setcounter{figure}{7}
\begin{figure*}[b]
    \includegraphics[width=\linewidth]{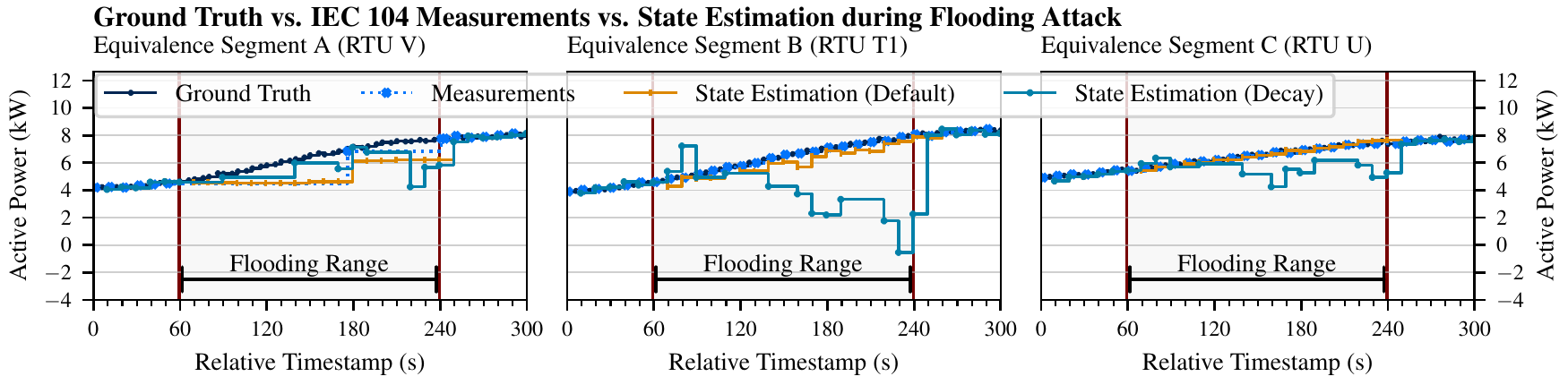}
    \caption{The flooding attack only affects the \acp{rtu} in equivalence segment $A$.
    During the attack, the majority of measurements of these \acp{rtu} do not reach the \ac{mtu}, leading to a loss of visibility and divergence of real and observed states.
    Since \num{51} \acp{rtu} communicating over the flooded links are affected, only \SI{\approx58}{\percent} of the measurements reach the \ac{mtu}.
    As a result, even the state estimations cannot fully compensate for the lack of information.}
    \label{fig:flooding-results}
\end{figure*}

The attackers physically destroy the equipment at \SI{\approx50}{\second}, disconnecting the entire substation from the grid and causing a locally constrained blackout.
Consequently, the current over the power line connected to this substation immediately drops to \SI{0}{\ampere}.
Furthermore, this attack interrupts the communication between the substation and the control center, mainly affecting the periodic measurements of the corresponding \ac{rtu}~$D$.
Therefore, the \ac{mtu} attempts to re-establish the connection after the TCP connection times out, indicated by the repeated TCP SYN packets in the communication plot.
Since the \ac{rtu} is disconnected from the network, the responsible router sends ICMP packets to the \ac{mtu}, stating its unreachability.
Since the measurements of an adjacent substation (\ac{rtu}~$N$) cover the impacted power line, the control center can localize the defects.
As shown in Fig.~\ref{fig:physical-attack}, they are also covered by the state estimation.

Through conducting this attack with \NAME{}, we can understand the locally limited impact of physical attacks on power grids.
Furthermore, this simulation illustrates that operators have various means on the power grid and communication side to detect such attacks, as their immediate impact is comparable to a physical outage.
However, attackers might also exploit their access to target a specific remote device, as presented in the following.

\subsection{Syntactic Attacks on Grid Communication}
\label{sec:attacks:syntactical}
Increasing the complexity of attacks, we continue with assessing the potential harm of syntactic attacks against the \ac{ict} network, aiming at a \ac{dos} condition: a TCP SYN flooding~\cite{eddy2007tcp} attack against a network branch and \ac{arp} spoofing~\cite{whalen2001introduction,ramachandran2005detecting} as a more precise attack measure.

\subsubsection{TCP SYN Flooding}

\setcounter{figure}{6}
\begin{figure}
    \centering
    \includegraphics[width=\linewidth]{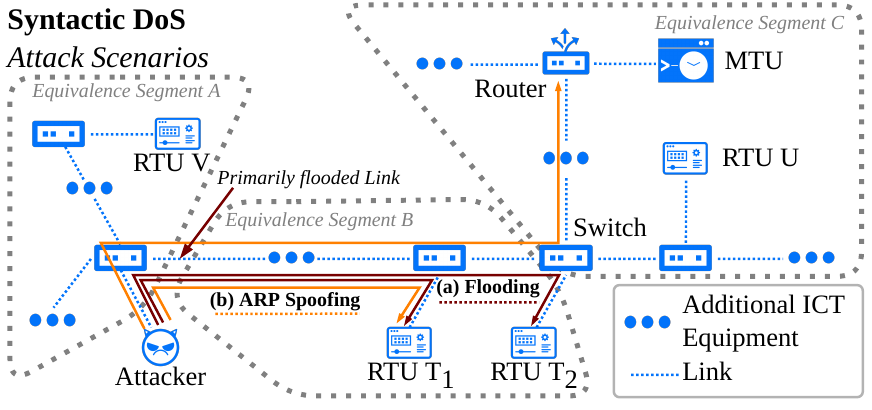}
    \caption{For both syntactic attacks, we attach an attack host via a \SI{1}{\giga\bit} link to a switch in the network.
    During the flooding attack~(a), attackers flood \acp{rtu} $T_1$ and $T_2$, mainly saturating the indicated link.
    During \ac{arp} spoofing~(b), attackers poison the \ac{arp} caches of \ac{rtu} $T_1$ and the router.}
    \label{fig:dos}
\end{figure}

Depending on the networking equipment and the targeted \acp{rtu}, such an attack might saturate the network's throughput capabilities or exhaust the resources at the targeted host by creating thousands of TCP contexts.
As a result, connections might get delayed, leading to higher packet loss and even collapsing TCP connections due to excessive delays.
To comprehensively understand these effects, we divide the considered network into equivalence segments following the network configuration and topology, as shown in Fig.~\ref{fig:dos}.
We expect a comparable impact on the devices belonging to the same segment.
With this approach, we can subsequently analyze the TCP SYN flooding impact on each segment's representative device to understand this impact.
We use \texttt{hping3} with three processes for each target, flooding the target with randomized source addresses and a TCP payload of \SI{1400}{\byte}.
The attack is executed from a single host, attached to a switch via a \SI{1}{\giga\bit} link.

Fig.~\ref{fig:flooding-results} shows the impact of this attack on the ability to reliably communicate measurements of the grid state for the different equivalence segments.
Due to the attackers' position in the network, the IEC~104 communication of the two targeted \acp{rtu} $T_1$ and $T_2$ is not notably affected by the attack since the primarily flooded link is not on the path between either \acp{rtu} $T_1$ and $T_2$, and the \ac{mtu} (equivalence segment $B$).
For similar reasons, \ac{rtu} $U$'s communication is not impaired, as holds for all \acp{rtu} in equivalence segment $C$.
However, since the communication of \ac{rtu} $V$ (and all other \acp{rtu} in equivalence segment $A$, \num{51} \acp{rtu}) entirely depends on the saturated link segment, it loses connectivity to the \ac{mtu} during the attack, resulting in the absence of measurements and a loss of visibility for the control center.

In particular, the attack leads to a loss of periodic measurements from \num{51} of the \num{119} \acp{rtu}, further affecting the grid state estimation.
As shown in Fig.~\ref{fig:flooding-results}, a loss of up to \SI{42}{\percent} of measurements heavily impacts the state estimation accuracy, as both applied estimation modes exhibit significant inaccuracies during the attack.
Here, assuming formerly reported measurements to remain valid when periodic transmissions are missing (\textit{\ac{se} Default}) results in the estimation to stay close to these outdated measurements.
In turn, dropping outdated measurements (\textit{\ac{se} Decay}) occasionally achieves a more precise estimate of the actual grid state but also exhibits large deviations.

In summary, we could reveal the profound linkage between the \ac{ict} and the power grid, where even a simple flooding attack significantly impacts the visibility and controllability of the power grid.
Moreover, the impact mainly depends on the attackers' location within the network and the timing of the attack, as the sole loss of visibility does not necessarily induce a critical grid state.
However, in the presence of significant changes in power consumption, the lack of visibility might destabilize the grid, trigger wrong control action decisions or prevent their realization.
Still, such attacks are rather evident to grid operators.

\subsubsection{ARP Spoofing}
The effectiveness of TCP SYN flooding strongly depends on the attackers' attachment point within the network.
In contrast, \ac{arp} spoofing~\cite{ramachandran2005detecting} is a more targeted attack strategy for \ac{dos}, potentially allowing attackers to intercept and interrupt IP-based communication within a subnet.
By sending forged \ac{arp} replies, attackers can poison a victim's \ac{arp} cache and manipulate the mapping of IP addresses to MAC addresses.
Essentially, the attackers announce themselves as the host with the IP addresses of their victims, causing corresponding packets to be forwarded to the attack host instead of the original destination.
Consequently, they can eavesdrop on the communication and especially drop arriving packets.

\setcounter{figure}{8}
\begin{figure}
    \includegraphics[width=\linewidth]{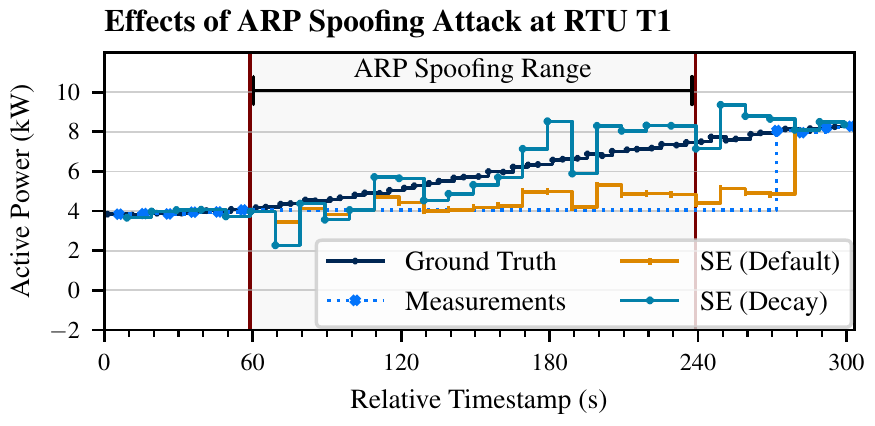}
    \caption{The \ac{arp} spoofing attack intercepts all traffic between the \ac{mtu} and \ac{rtu} $T_1$.
    As a result, the connection drops and no measurements reach the \ac{mtu}.
    After the attack, it takes \SI{\approx30}{\second} for the \ac{arp} cache to recover and re-establishing the connection.
    Although the attack only affects a single host, the state estimation accuracy for measurements of \ac{rtu} $T_1$ decreases.}
    \label{fig:arp-spoofing}
\end{figure}

For the spoofing attack, we use the same configuration as in the TCP SYN flooding attack scenario (cf.~Fig.~\ref{fig:dos}).
We send spoofed \ac{arp} packets to the router and \ac{rtu} $T_1$ using \texttt{arpspoof}.
After poisoning the \ac{arp} caches, the traffic between the two hosts is redirected to the attack host, which drops all traffic and thus induces a \ac{dos} condition.
Thus, the TCP connection between the \ac{mtu} and \ac{rtu} $T_1$ times out, and no more measurements from the \ac{rtu} arrive at the control center.
As the results in Fig.~\ref{fig:arp-spoofing} indicate, even the loss of a single information source results in a decreased accuracy of the state estimation for the affected grid components, especially for the default \ac{se} behavior.
After the \ac{arp} spoofing attack stops, the \ac{arp} cache takes another \SI{\approx30}{\second} to recover from the poisoning, i.e., the attack effects last longer than the actual attack.

Compared to TCP SYN flooding attacks, \ac{arp} spoofing allows for more precise target selection since only the targeted \ac{rtu} $T_1$ is affected by the attack.
However, as the combined results of \NAME{} show, even losing a single \ac{rtu} affects the visibility within the grid and might even interfere with control commands.
Still, such attacks can be easily detected, as missing measurements clearly indicate an anomaly (defects or attacks).

\subsection{Comprehensive Semantic Attacks}
\label{sec:attacks:semantic}
While syntactic attacks (cf. \S\ref{sec:attacks:syntactical}) might achieve a loss of visibility and control within the power grid, they do not actively interfere with the grid operation.
Further, they are easily detectable and preventable.
Thus, we shift our focus from syntactic attacks to more sophisticated domain-specific semantic attacks.
Here, attackers actively manipulate and inject commands and monitoring information by utilizing knowledge about used protocols, the grid's configuration, and the semantic meaning of network communication.
In the following, we first analyze Industroyer~\cite{cherepanov2017win32}, a real-world semantic attack that led to an actual blackout, before we study a stealthy false data injection attack.

\subsubsection{The \emph{Industroyer} Circuit Breaker Attack}
The Industroyer family~\cite{cherepanov2017win32}, which, among other things, targets IEC~104-based networks by sending control commands to \acp{rtu}, is a real-world example of such a semantic interference strategy.
During the Ukrainian power grid attack in 2016~\cite{mcfail2022detection}, the Industroyer malware successfully cut power for a fifth of the capital Kyiv.
In the more recent Ukrainian power grid attacks in April 2022, its successor, the so-called \emph{Industroyer2}~\cite{industroyer2blogpost}, was (unsuccessfully) used by attackers aiming to again induce a blackout~\cite{industroyer2blogpost}.
Industroyer attempts to connect to one or multiple pre-configured IP addresses as an IEC~104 client for issuing control commands.
Thus, the attackers need prior insider knowledge on the devices (\acp{rtu}) w.r.t. their IP addresses and IEC~104 configuration, i.e., the malware configuration further includes the common address (COA) and associated \acp{ioa} for each configured \ac{rtu}~\cite{cherepanov2017win32,industroyer2blogpost}.
By issuing control commands to open or close circuit breakers, Industroyer disconnects or connects the respective grid parts from the power supply.

\begin{figure}
    \includegraphics[width=\linewidth]{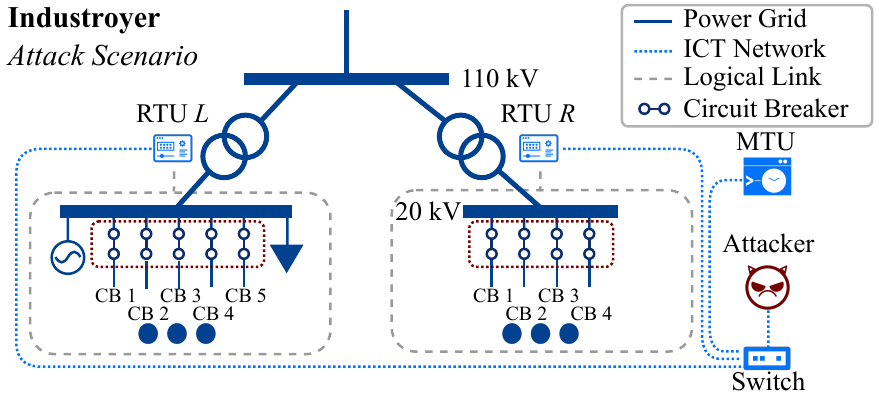}
    \caption{The Industroyer malware connects to two \acp{rtu} placed at the central substation of the Simbench reference grid.
    During the attack, the malware issues single-point commands to open all circuit breakers managed by these \acp{rtu} to disrupt the power supply of the grid.}
    \label{fig:industroyer-scenario}
\end{figure}

We implement an attack reproducing the behavior of the Industroyer malware in \NAME{} to analyze its potential impact on the power grid.
As we depict in Fig.~\ref{fig:industroyer-scenario}, we deploy an attack host configured to connect to two \acp{rtu} responsible for the central substation of the grid.
Immediately after establishing these connections, the attackers repeatedly issue IEC~104 single-point commands for two minutes, causing the circuit breakers to open, effectively disconnecting most of the medium voltage grid from the high voltage grid.
Closely following the implementation of the Industroyer malware, we use a dedicated thread per \ac{rtu}, each issuing control commands in a \SI{3}{\second} interval~\cite{industroyer2analysis}.

\begin{figure}
    \vspace{-6pt} %
    \includegraphics[width=\linewidth]{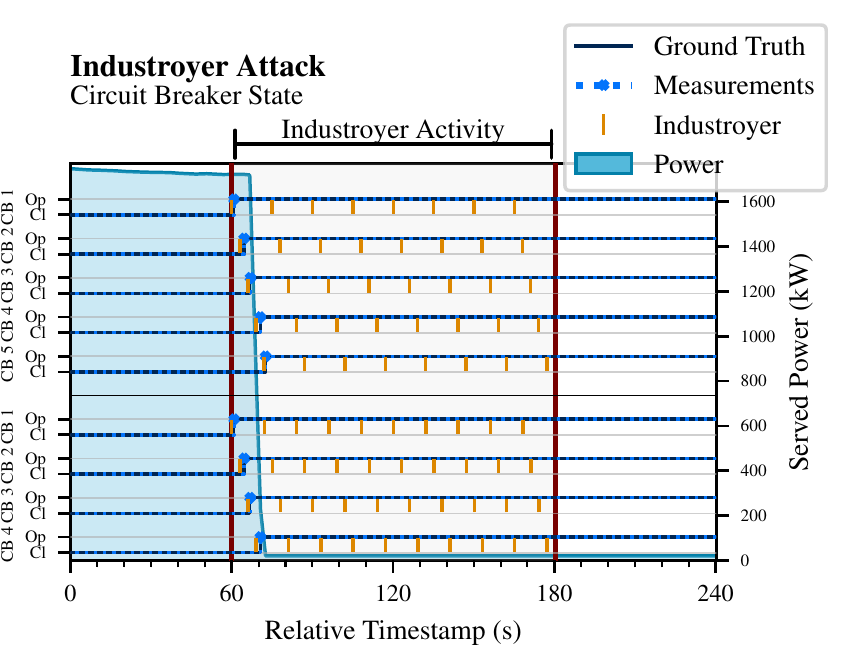}
    \caption{As soon as the Industroyer malware connects to both \acp{rtu}, it repeatedly issues control commands to open (Op) circuit breakers (CB).
    This ensures that even if the \ac{mtu} issues commands to close (Cl) the circuit breakers, they are opened again.
    The attack results in more than \SI{1.6}{\mega\watt} of load being disconnected from the grid, causing a blackout.
    }
    \label{fig:industroyer-plot}
\end{figure}

In Fig.~\ref{fig:industroyer-plot}, we visualize the command injections, their effects on the circuit breakers, and the total power load.
When the attack starts, the attackers issue control commands to open the pre-configured circuit breakers.
After \SI{12}{\second}, a command for each circuit breaker has been issued, disconnecting more than \SI{1.6}{\mega\watt} of load from the grid.
Essentially, this creates a large-scale blackout with little potential for effective countermeasures by the operator.
Even after the attack stops, restoring the grid's full operation can take several hours to days, as equipment might be damaged, and the grid requires a gradual restart~\cite{patsakis2018optimal}.

As Industroyer does not rely on \ac{arp} spoofing, preventing it requires further measures besides, e.g., static \ac{arp} tables.
Even the detection of unauthorized devices would not suffice to detect and prevent this attack since, in the real-world attack, the malware was deployed on hosts controlled by the grid operator~\cite{industroyer2blogpost}.
However, %
a stringent whitelist of allowed IEC~104 clients and/or message authentication can reduce the chance of a successful attack.
We exemplary verified the impact of a whitelisting approach:
Configuring \acp{rtu} to reject IEC~104 connections from hosts different from the \ac{mtu} renders the current Industroyer attack unsuccessful (thus attackers, e.g., would additionally need to spoof the IP address of the MTU).

Our results obtained using \NAME{} are twofold:
First, they underline the significant security flaws of real-world power grids that enable such attacks.
With only limited knowledge about the power grid and its network configuration, attackers are able to cause significant physical damage and induce a blackout. %
Second, they demonstrate that implementing countermeasures, e.g., whitelisting IEC~104 clients, effectively hamper attack success.

\subsubsection{False Data Injection}
\label{subsec:attack:fdi}
\begin{figure}
    \includegraphics[width=\linewidth]{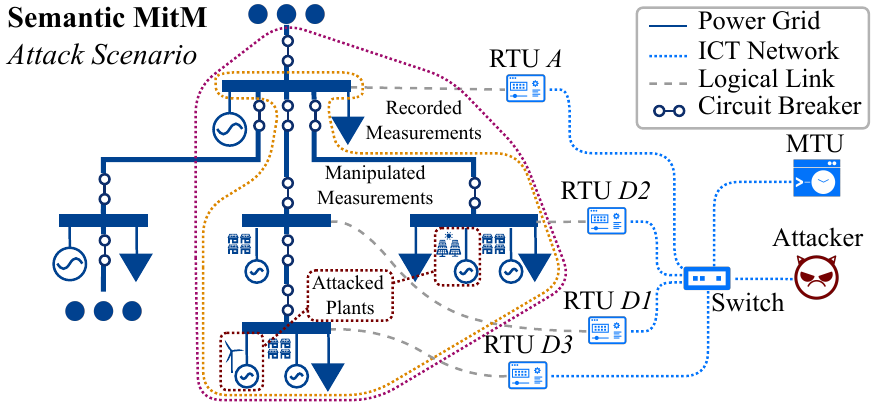}
    \caption{For the false injection, the attackers place themselves as an \ac{mitm} between the \ac{mtu} and four \acp{rtu}.
    After an initial recording phase, the attackers transparently inject commands for deactivating the power infeed of two renewable power plants.
    Subsequent measurement transmissions are manipulated to reflect the original grid state.}
    \label{fig:mitm-scenario}
\end{figure}

Despite the severe consequences of the Industroyer attack, its effects are immediately visible to the grid operator.
Therefore, we now assume sophisticated attackers with advanced knowledge, targeting to transparently manipulate measurements sent to the control center to obfuscate the attack's effects in a future-oriented scenario with numerous renewable power sources attached to the \ac{ict} network.
Similar to the previous attacks, we deploy an attack host, as shown in Fig.~\ref{fig:mitm-scenario}, which establishes itself as an \ac{mitm} between the \ac{mtu} and multiple \acp{rtu} using \ac{arp} spoofing.
Alternatively, attackers might also physically %
reconnect networking cables via the attack host.
To evaluate the potential cascading effects of the attack, we deploy virtual safety devices at each circuit breaker, monitoring voltage, and current limits and disconnecting parts of the grid if these limits are exceeded.

\begin{figure*}
    \includegraphics[width=\linewidth]{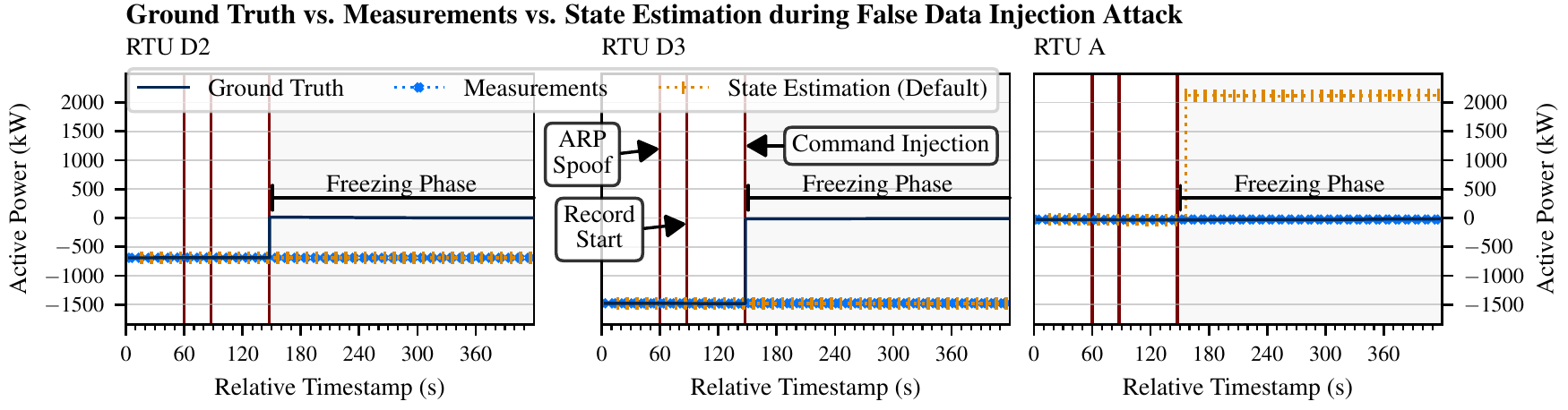}
    \caption{During the false data injection attack, the attackers establish themselves as an \ac{mitm} and record monitoring information sent by multiple \acp{rtu} for \SI{60}{\second}.
    Afterwards, they inject two control commands to disable the power infeed of two renewable power plants controlled by \acp{rtu} $D2$ and $D3$. %
    However, to obfuscate the effects of their attacks, the attackers replace subsequent measurements with values matching the grid state before the command injection.
    As a consequence, the actual grid state, reported measurements, and the state estimation (SE) results differ significantly.
    }
    \label{fig:mitm-plot}
\end{figure*}

As an \ac{mitm}, the attack host records all measurements and control commands for \SI{60}{\second}.
Afterward, it injects two control commands to shut down the power infeed of two renewable plants at \acp{rtu} $D2$ and $D3$, as indicated in Fig.~\ref{fig:mitm-scenario}, and suppresses the respective command confirmations (to hide the command injection from the control center).
For each injection, manipulation, or suppression of packets, the attackers adjust the sequence and acknowledgment numbers for the TCP and IEC~104 connections.
Hence, the original communication partners are entirely unaware of the intercepted communication.
To conceal the physical impact of the injected commands, the attackers replace subsequently communicated measurements of the targeted \acp{rtu} with those matching the trend of the initial recording phase (\emph{freezing}).
Further, they suppress any control commands sent from the control center (while still acknowledging them), yielding the operator without control over the attacked \acp{rtu}.

As shown in Fig.~\ref{fig:mitm-plot}, the active power infeed at the buses controlled and monitored by \acp{rtu} $D2$ and $D3$ drops close to \SI{0}{\kilo\watt} after the command injection at \SI{150}{\second}.
However, the manipulated measurements still report an active power injection of \SI{691}{\kilo\watt} and \SI{1.47}{\mega\watt}, respectively.
Due to the inconsistencies between the measurements from the attacked \acp{rtu} and the remaining \acp{rtu}, the control center's state estimation miscalculates the grid state significantly, incorrectly suggesting an increased power drain of more than \SI{2}{\mega\watt} at the bus at \ac{rtu} $A$.

Besides hindering the localization of the actual event leading to these deviations and complicating the detection of the attack along with potential countermeasures by the grid operator, this attack might even provoke improper reactions by the grid operator, e.g., targeted load shedding at the bus controlled by \ac{rtu} $A$.
Depending on the attackers' technique for establishing themselves as an \ac{mitm}, stopping the ongoing attack might require physical interaction of the grid operator and time-consuming reconnections of power grid assets.
The combination of active interference with the grid operation by hiddenly injecting control commands and the subsequent false data injection for obfuscating the commands' existence and effects thus emerges as a particularly critical attack scenario.

To summarize, our evaluation using \NAME{} illustrates the inseparable connection between \ac{ict} and physical processes in power grids.
Further, we see that attackers can exploit knowledge of the power grid's structure and operation to amplify the immediate effects of attacks.

Overall, our analysis of the impact of various cyberattacks on power grids using \NAME{} shows the importance of comprehensively considering effects on the communication and energy side of power grids to fully understand the attack capabilities, thus paving the way for adequately securing tomorrow's power grids.

\section{Discussion and Future Research}%
\label{sec:discussion}

Using \NAME{}, we reproduced and analyzed different attacks from various attack classes (cf.\ \S\ref{sec:security:attacks}) where the attackers increasingly require more knowledge about the power grid, correlating to the criticality of the attack's impact.
In the following, we discuss the insights gained during our evaluation of the impact of cyberattacks (cf.\ \S\ref{sec:attacks}) and identify further research potential. %

As the least sophisticated attacks, \emph{physical attacks}, e.g., destroying equipment, typically only have a locally limited impact on the power grid operation due to the redundant structure of such grids.
Here, \NAME{} enables an impact assessment of outages of different components, thus identifying which components require special protection.
Moreover, it spurs improving the detection of outages, regardless of whether they result from a cyberattack or a purely technical failure, e.g., by identifying distinct artifacts of such outages w.r.t. to communication behavior.

In turn, \emph{syntactic attacks} can impede the visibility of the grid operator. %
Moreover, we also understand that syntactic \ac{dos} attacks do not actively change the grid state, and the grid remains operational as long as no control commands are required.
While power grids traditionally only occasionally required active control~\cite{monti2010power}, increasing power demands and renewable power sources nowadays lead to a higher necessity for active control commands.
Thus, such modern power grids are more affected by \ac{dos} attacks against the communication network.
Nevertheless, albeit rarely performed in practice, such attacks are preventable and easily detectable, e.g., with static \ac{arp} tables, switch port authentication or deactivation, rate limiting, and \acp{ids}.
Here, \NAME{} provides a safe environment to test such security measures before rolling them out.

Most importantly, we studied complex \emph{semantic attacks}, %
illustrating how attackers can physically damage the power grid by (remotely) exploiting susceptible \ac{ict}.
In particular, our results show that false data injection attacks involve a high risk for grid operators since they actively manipulate the grid's state while also obfuscating the attack's impact, leading to delayed reactions.
In the extreme case, a sophisticated attack can cut off parts of the grid or permanently damage physical assets while simultaneously faking normal grid conditions to the grid operator.
Besides statically freezing the current grid state, such an attack could also react to changes in other regions of the grid and influence the power flow, making it virtually impossible to detect the attack (both on the \ac{ict} and energy side) based on received measurements.

While we primarily designed and demonstrated \NAME{} for analyzing the impact of cyberattacks, our discussion indicates enormous \emph{further potential} for \NAME{} to strengthen cybersecurity in power grids.
As evident from our analysis and discussion, there is an inherent need to roll out measures to prevent (e.g., encryption and authentication of commands and measurements) and detect (e.g., intrusion detection systems) such cyberattacks.

However, power grid operators are typically reluctant to roll out any changes to their infrastructure as they fear negative impacts on grid operation and often feel overwhelmed with selecting, prioritizing, and adapting security measures.
For example, power grid operators have been reluctant to introduce comprehensive authentication and encryption since intensive computations for long-living, resource-constrained devices contradict the power grid's stringent availability and latency requirements~\cite{esiner2019fpro,hiller2018secure,henze2017distributed}.
Likewise, when considering intrusion detection systems, operators face the challenge of reliably locating an attack's origin within geographically widespread power grid networks~\cite{bace2001intrusion} without extensive sensor placements~\cite{chen2010optimising}.
Here, \NAME{} can serve as a safe environment to try out and evaluate such approaches for the prevention and detection of attacks, e.g., to ensure that security measures do not impact the grid's availability or to assist in identifying optimal sensor placement for detection approaches.
Indeed, \NAME{} has already been applied to evaluate an intrusion detection approach based on discrete-time Markov chains~\cite{wolsing2022ipal}.
In ongoing work, \NAME{} provides the foundation to study how network intrusion detection in power grids can be enhanced with automated facility monitoring~\cite{serror2022inside}.

\textbf{Ethical Considerations.}
We acknowledge potential risks arising from our work, especially for power grid operators.
However, we are convinced that the advantages of evaluating cyberattacks for understanding their impact and deriving suitable countermeasures outweigh the risks.
We further adopt multiple measures to minimize such risks:
First, we do not develop entirely new cyberattacks against power grids but focus on existing ones.
Second, we refrain from publishing any implementation of the conducted attacks to prevent misuse of our research.
This is in line with the judgment of domain experts from a national cybersecurity agency and multiple grid operators with whom we discussed the performed attacks.

\section{Conclusion}%
\label{sec:conclusion}

This paper addresses the challenge of comprehensively researching cybersecurity within power grids as a cyber-physical system.
Our related work analysis reveals that existing research environments do not sufficiently cover the requirements for such a comprehensive methodology.
Therefore, we introduce \NAME{}, an open-source research environment for studying sophisticated cyberattacks against power grids of realistic sizes and, primarily, their impact on communication and physical processes.
We validated \NAME{}'s accuracy against a physical testbed, showed its scalability using benchmarking scenarios and reference power grids, and expounded its flexibility and suitability for cybersecurity research.

Accordingly, we used \NAME{} to recreate and analyze different attacks of increasing complexity against a realistic reference distribution grid~\cite{meinecke2020simbench}, including the sophisticated Industroyer attack, which was used to attack Ukrainian power grids~\cite{industroyer2blogpost}.
Our analysis of the impact of different attacks reveals that with increasing technical knowledge of the grid, attackers can shift from a locally limited attack impact to an active disruption of the power grid's operation and even cause widespread blackouts.

While such attacks can already be devastating, obfuscating their effects by manipulating measurements and interfering with countermeasures further aggravates their impact.
\NAME{} provides the necessary means to researchers and grid operators for analyzing and comprehending the combined impact of such advanced cyberattacks on the communication and physical side of power grids.
Moving forward in securing power grids based on this knowledge, \NAME{} offers a safe and risk-free environment to test countermeasures, validate their effectiveness, and ensure the absence of unintended side effects on reliable power grid operation.

\section*{Acknowledgments}
We thank Martin Unkel for his indispensable work on and support for the implementation of the C\texttt{++} Python bindings\footnote{\url{https://github.com/fraunhofer-fit-coop/104-connector-python}} for the IEC~60870-5-104 protocol.
These bindings represent a vital component used in \NAME{}'s \ac{rtu} and \ac{mtu} reference implementations.

\bibliographystyle{plain}
\bibliography{paper}

\end{document}